\documentclass[runningheads]{llncs}
\usepackage[T1]{fontenc}

\usepackage{graphicx}

\usepackage{mathtools}
\usepackage{amssymb}

\begin{document}
\title{EMG subspace alignment and visualization for cross-subject hand gesture classification}
\titlerunning{Cross-subject EMG subspace alignment and visualization}
\author{Martin~Colot\inst{1} \and
Cédric~Simar\inst{1, 2} \and
Mathieu~Petieau\inst{2} \and Ana~Maria~Cebolla~Alvarez\inst{2} \and Guy~Cheron\inst{2} \and Gianluca~Bontempi\inst{1}}
\authorrunning{M. Colot et al.}

\institute{Machine Learning Group - ULB, Belgium\and
Laboratory of Neurophysiology and Movement Biomechanics (LNMB)
and ULB-Neuroscience Institut (UNI) - ULB, Belgium}
\maketitle            
\begin{abstract}
Electromyograms (EMG)-based hand gesture recognition systems are a promising technology for human/machine interfaces. However, one of their main limitations is the long calibration time that is typically required to handle new users. The paper discusses and analyses the challenge of cross-subject generalization thanks to an original dataset containing the EMG signals of 14 human subjects during hand gestures. The experimental results show that, though an accurate generalization based on pooling multiple subjects is hardly achievable, it is possible to improve the cross-subject estimation by identifying a robust low-dimensional subspace for multiple subjects and aligning it to a target subject. A visualization of the subspace enables us to provide insights for the improvement of cross-subject generalization with EMG signals.

\end{abstract}
\section{Introduction}

The recognition of hand gestures from electromyographic (EMG) signals is important for many brain-computer interfaces (BCI) applications, such as robotic hand prostheses, video game control, or sign language recognition. 
The muscle activity is recorded through surface electrodes, placed on the skin, and processed by a machine learning model to classify instantaneous hand postures.
These models have become very accurate when considering a single subject and session~\cite{ref:literatureReview}. However, they often lack generalization across subjects. 
%
A major difficulty when working with EMG signals is that they are highly person-specific~\cite{hoshino2022comparing,gu2020cross} due to intrinsic differences in anatomical and physiological characteristics.
Also, small shifts in the location of the electrodes can significantly affect the recorded signal.
In order to deal with real-world settings, EMG-based systems must tackle the \emph{cross-subject} issue, i.e., be able to generalize to other users (targets) than the ones used for training (sources). 
This is necessary since long phases of labeled data collection are not acceptable in real settings, and may even be impossible for impaired users.
We tackle this issue as an unsupervised domain adaptation (UDA) problem where labeled samples from multiple source subjects are available, and a model has to classify unlabelled samples from a target subject. 
In a real-world implementation, the model would have an initial low efficiency with a new user. Then, as new unlabelled samples are collected, the model parameters would regularly be updated through UDA to improve the estimation.

In this paper, we focus on bridging the gap between intra-subject (training and testing on the same person) and cross-subject classification by analyzing an original EMG dataset where 14 participants perform 4 simple hand postures.
We start by considering several dimensionality reduction strategies to better understand the dissimilarities that occur in the dataset and find a simpler representation of the samples.
We then show that a common low-dimensional subspace can help to perform unsupervised domain adaptation, improving the classification accuracy on a target subject.
The paper presents the following results:
i) an assessment of hand gesture classification from EMG in intra-subject and cross-subject configurations in the original signal space,
ii) the definition and visualization of a common low-dimensional subspace, and
iii) the assessment of subspace alignment domain adaptation with a leave-one-subject-out strategy.

\section{Related Work \label{sec:relatedWork}}
%
Classification of hand gestures from EMG signals has attracted large attention in machine learning~\cite{ref:literatureReview}. However, since the number of classes, hand gestures, and data acquisition are very heterogeneous over different studies, it is hard to define a state-of-the-art baseline accuracy. Nevertheless, it appears that, when few classes are considered, the intra-subject classification accuracy is typically above 90\%~\cite{ref:literatureReview}.
%
EMG signals are usually classified using either a shallow machine-learning model with handcrafted features or a deep CNN with the raw signal ~\cite{hoshino2022comparing,fajardo2021emg}. 
Recently, CNN has become the primary choice for EMG classification. However, with few signs to recognize, it is often possible to obtain similar results with a shallow neural network or handcrafted features~\cite{fajardo2021emg}.

Cross-subject classification usually involves training a single classifier on pooled samples from many source subjects, with the aim of obtaining enough generalization to handle new subjects.
However,~\cite{hoshino2022comparing} stresses that, without fine-tuning for the target subject, the accuracy stays low even with many sources subjects.
%
We refer to unsupervised domain adaptation as transfer learning methods that use unlabeled samples from a target domain that is different but related to the source domain. 
Cross-subject EMG classification is typically a multi-sources domain adaptation problem as we can access labeled samples from a set of source subjects and adapt the model to a target subject using only unlabeled samples.
This problem has recently attracted attention for classifying high-density EMG (HD-EMG) with the implementations of deep neural network-based methods.
In~\cite{ref:domainAdaptationEMG1}, the AdaBN method is used to adapt a deep convolutional neural network. In the cross-subject configuration, their method improves the accuracy from 39\% to 55.3\% with 10 subjects and 8 hand gestures. 
In~\cite{ref:domainAdaptationEMG2}, a sampling strategy that aligns the marginal distributions of the sources and target domains is introduced. They show that it adapts the model better than AdaBN, up to closing the gap between intra-subjet and cross-subject with HD-EMG.

\section{The experimental setting}
Our experiments are conducted with an original dataset of 14 participants. It contains 400ms non-overlapping windows of EMG during the hold of 4 different finger postures. The data acquisition protocol is detailed in appendix A\footnote{This dataset is expected to be published shortly in another paper.}~\cite{ref:masterThesis}.
To classify those samples, we use engineered features that have led to good EMG classification in related studies~\cite{fajardo2021emg}.
Those are the mean absolute value, root mean square, waveform length, zero crossing, Wilson amplitude, maximum absolute amplitude, and integral. We compute them on each of the 8 EMG channels, which gives us a feature vector of size $56$.

\section{Baseline classifiers}
For the intra-subject baseline, we train one classifier for each subject; we take 90\% of the samples for training and 10\% for testing in a 10-fold cross-validation. We then compute the average classification accuracy.
For cross-subject configurations, leave-one-subject-out cross-validation is used. We pool the samples from 13 subjects to create the training set and use those from the remaining subject for testing.
All our experiments are conducted using similar classification models. We use a multi-layer perceptron neural network with one hidden layer and a logistic activation function. We keep 90\% of the training set for fitting the model and 10\% as the validation set. The model is trained until convergence of the classification accuracy on the validation set, with a maximum of 1000 epochs. The hidden layer is set to contain 100 nodes in the intra-subject configurations and only 10 in the cross-subject configurations. This is set to reduce the effect of overfitting in cross-subject configurations.

The baseline classification reaches 93.1\% accuracy for intra-subject and 69.8\% for cross-subject.
While this result highlights the \emph{cross-subject} issue, it also assesses the quality of our model as its performance is similar to the state-of-the-art (typically above 90\% for intra-subject models).
For the following tests, we consider that the model has access to all the unlabelled target samples at training time and that those samples contain only EMG samples from the 4 recognized classes. In a real-world setting, the model should be able to discriminate between samples that correspond to a specific gesture class or no intended gesture. We simplify the problem to concentrate on highlighting the cross-subject issue.

\section{Dimensionality reduction for cross-subject learning \label{sec:dim_reduction}}

\begin{figure}
\begin{center}
\includegraphics[width=.9\textwidth]{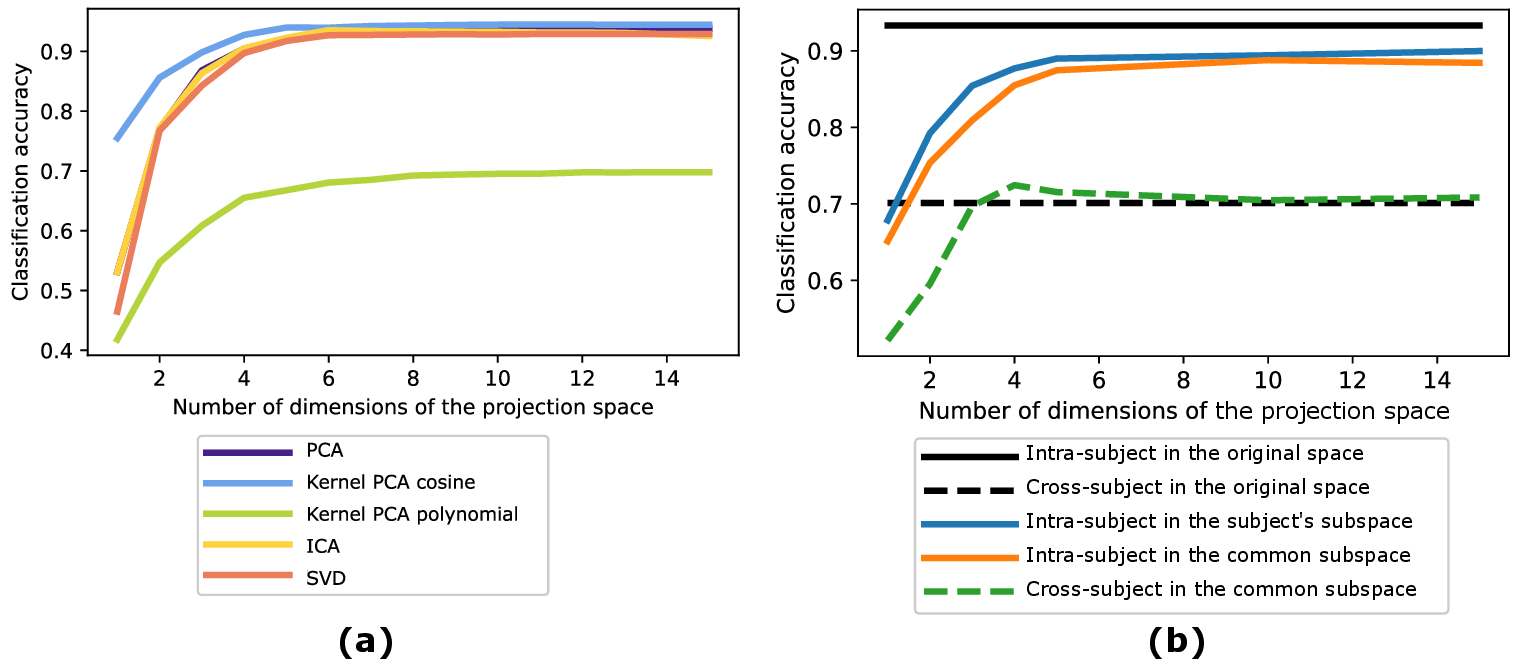}
\caption{\textbf{(a)} Intra-subject classification accuracy for different dimensionality reduction algorithms and an increasing number of dimensions of the projection space. \textbf{(b)} Classification accuracy for an increasing number of dimensions of the KPCA subspace.} \label{fig_subspace}
\end{center}
\end{figure}

This section explores the role of dimensionality reduction in improving cross-subject generalization. The rationale of the analysis is that the poor accuracy of the cross-subject classification observed in the previous section might be improved by finding a subspace where the intra-subject conditional distributions are more similar.
We consider the following dimensionality reduction strategies: Principal Component Analysis (PCA), Kernel PCA (KPCA)  with cosine and polynomial kernel, independent component analysis (ICA), and singular value decomposition (SVD). 
The plot in Figure \ref{fig_subspace}.a reports the classification accuracy in intra-subject configurations, where these algorithms are fitted on each subject separately.
It appears that KPCA with a cosine kernel offers the best data representation with few dimensions.
We evaluate the quality of this subspace for cross-subject estimation. Considering a leave-one-subject-out strategy, we fit the projection on the pooled samples from all sources subjects to find a common subspace. We project all the samples in this subspace and compute the cross-subject accuracy. The results are shown in Figure \ref{fig_subspace}.b.
The main conclusions deriving from the conducted experiments are 
i) for a single subject, the EMG signals may be effectively embedded in a low-dimensional space so that a classification model keeps high accuracy,
ii) selecting a common subspace for all subjects does not help with the cross-subject issue.
It remains however an open question: do the subjects differ in terms of subspace or in terms of distributions, yet over the same subspace? 
In order to settle the question, we implement and assess a cross-subject dimensionality reduction step. The idea is to use all the subjects with one set aside (leave-one-subject-out) to derive a common low-dimension subspace and then test such subspace (source) on the remaining subject (target). 
Note that in this case, though the subspace has been obtained from the other subjects, the training set contains only the EMG signals of the target.

From the orange line in Figure \ref{fig_subspace}.b, it appears that the intra-subject accuracy in the common subspace falls just below the accuracy obtained in the target subject's specific subspace.
This suggests that the subspaces of different subjects are similar but not the same. The mapping between the features and labels must be different with each subject to explain the poor cross-subject accuracy.

\section{Visualization of the low-dimensional subspace}
\begin{figure}
    \centering
    \includegraphics[width=1\textwidth]{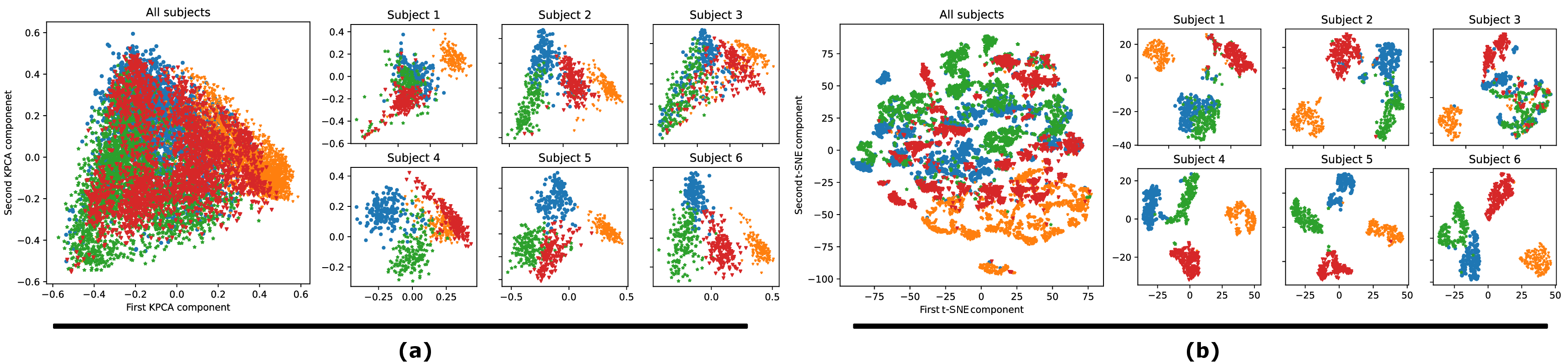}
    \caption{\textbf{(a)} KPCA projections of the samples from all subjects in a common subspace. \textbf{(b)} t-SNE visualization of the samples from all subjects in the common KPCA subspace}
    \label{Fig:projectionsPlot}
\end{figure}

We have shown in Section~\ref{sec:dim_reduction} that there is a low-dimensional KPCA subspace that is convenient for training a classifier for all the subjects. 
This section uses the first 2 components of such projection (computed from the entire dataset) to gain a visual insight into the distribution of both the pooled dataset (left side of Figure~\ref{Fig:projectionsPlot}.a) and the individual subjects (the right side of Figure~\ref{Fig:projectionsPlot}.a shows the first 6 subjects).
As we can see, when looking at one subject at a time, the class clusters are often separable, even with only 2 dimensions.
However, the shape and positions of those clusters are very subject-dependent.
The leftmost subfigure shows that once pooled all the subjects together, the class clusters are hardly separable. This explains the poor accuracy obtained with cross-subject training of the classification model.
To show that this effect stays in higher dimensions, we provide a t-SNE projection of the samples in Figure~\ref{Fig:projectionsPlot}.b. This projection is computed from the 10-dimensional common subspace as it provides good intra-subject classification.
This enables us to show that even when considering the complete low-dimensional subspace, the class clusters are easier to separate when looking at one subject at a time than when pooling the subjects together.

\section{Subspace alignment for domain adaptation}

\begin{figure}
\centering
\includegraphics[width=.9\textwidth]{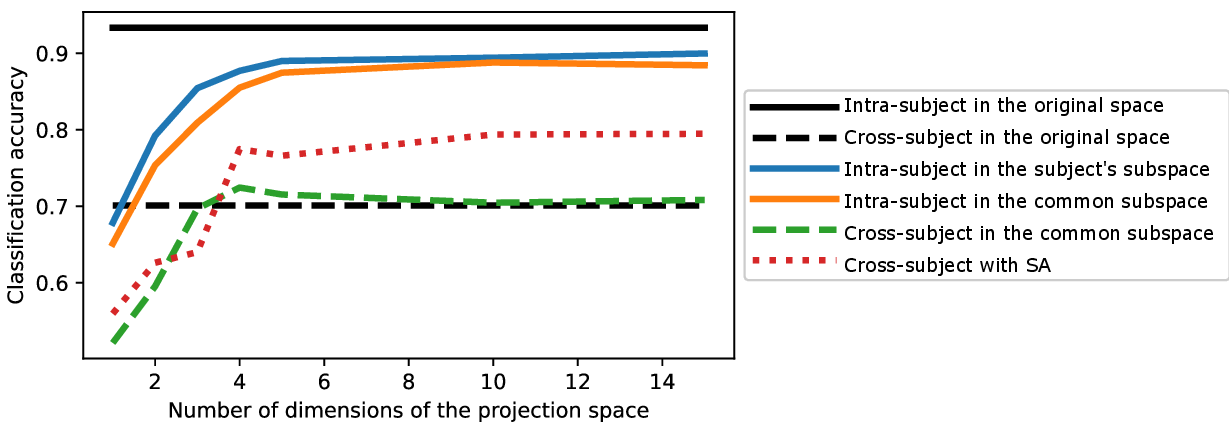}
\caption{Comparison of the classification accuracy obtained with KPCA SA and with the other strategies for an increasing number of dimensions of the KPCA subspace.} \label{fig_subspace_sa}
\end{figure}

\begin{figure}
\centering
\includegraphics[width=1\textwidth]{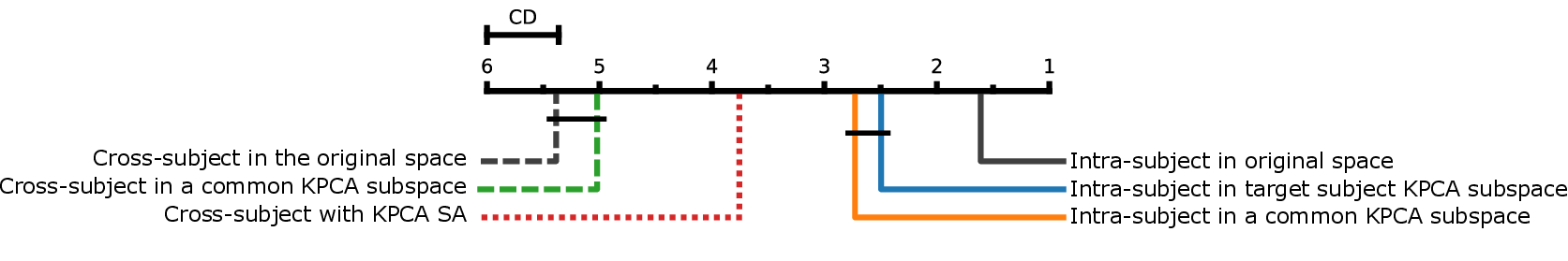}
\caption{Average rank of the different classifiers (the lower, the better). Classifiers that are not
significantly different are connected (at \textit{p = 0.05} found by a Nemenyi test~\cite{demvsar2006statistical}).} \label{fig_nemyeni}
\end{figure}

Subspace alignment (SA)~\cite{ref:SA} is a UDA method that projects the samples from two different but related domains into a single subspace using PCA projections of the two domains.
We suggest that applying this method to our engineered features will yield a common subspace, with a better correspondence between the source and target domains.
As in section \ref{sec:dim_reduction}, cosine KPCA gave better results than PCA; we adapted the simple SA to work with cosine kernel. As explained in~\cite{ref:cosine}, the cosine similarity function is equivalent to the linear kernel (used by PCA) if the data is L2-normalized.
Hence, applying L2-normalization on each subject enables us to use standard SA to align the same subspaces as the one obtained with cosine KPCA.
We implement an estimator, using all the source subjects as a single pooled domain, and apply SA to align the common KPCA subspace of those source subjects with the target subject's KPCA subspace.
We validate the alignment efficiency in leave-one-subject-out cross-validation.
The results in Figure \ref{fig_subspace_sa} show that this domain adaptation method obtains a better accuracy than the cross-subject's baseline.
The model converges to 79.5\% accuracy, obtained with only 10 dimensions of the KPCA subspace. The posthoc analysis of these results using a Nemenyi test~\cite{demvsar2006statistical} (given in Figure \ref{fig_nemyeni}) shows that this result is significantly better than the cross-subject baseline. However, it doesn't close the gap between cross-subject and intra-subject.

\begin{figure}
\centering
\includegraphics[width=1\textwidth]{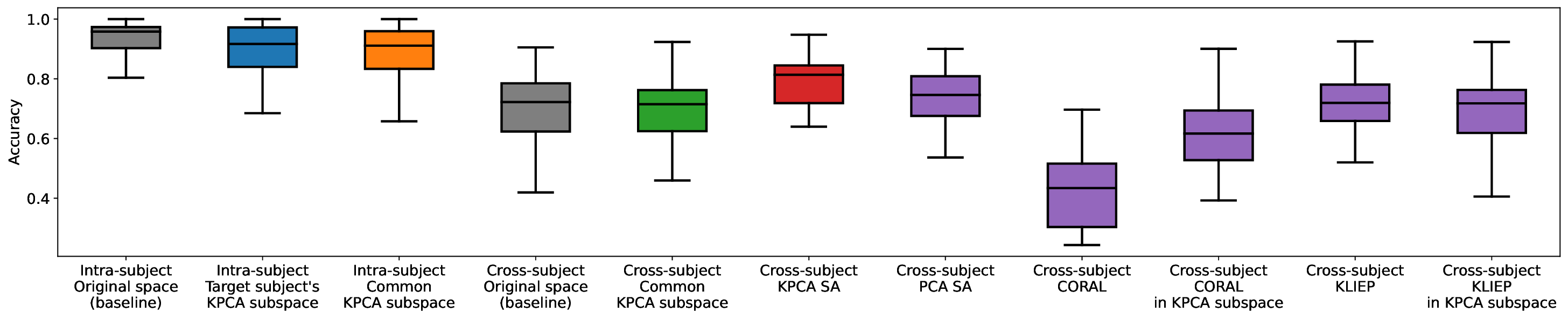}
\caption{Comparison of the results from the different models} \label{fig_results}
\end{figure}

We finish by comparing the KPCA SA with other UDA methods. 
First, we test the standard PCA SA by not applying L2-normalization to see if it has any impact on the model.
After that, we test the Correlation Alignment (CORAL) algorithm~\cite{ref:coral}. This method aligns the second-order statistics of the sources and target domains.
Finally, we introduce an instance-based domain adaptation algorithm. The Kullback-Leibler Importance Estimation Procedure (KLIEP)~\cite{ref:kliep} finds a weighing of the sources samples that minimizes the differences with the target domain distribution. This strategy does not affect the feature space, contrary to SA and CORAL. 
CORAL and KLIEP are tested both in the original feature space (with L2-normalization) and in the common KPCA subspace found on source subjects. We used 10 dimensions for the KPCA subspace as it gave the best results in figure \ref{fig_subspace_sa}.
The results in Figure \ref{fig_results} show that all of these models are less efficient than KPCA SA.

\section{Conclusion}
%
The literature on EMG-based hand gesture classification is full of highly accurate results as far as a single subject is concerned. The story changes when several subjects are pooled together.
This paper confirms the limitations of naive cross-subject approaches, which simply rely on pooling together the data of several subjects. At the same time, our results show that it is possible to find a common robust subspace in which classifiers can rapidly be tuned to different subjects. Though such a common subspace is low-dimensional, the associated classifiers are competitive with the ones fitted to specific individuals.
Moreover, we showed that this subspace helps improve the accuracy of cross-subject classifiers. Using SA to align the target's specific subspace to the common subspace found on sources, we improved the accuracy from 69.8\% to 79.5\%.

Such a result is encouraging, but it should be considered as a first step in the definition of a robust methodology to solve the cross-subject problem of learning from electrophysiological signals.
A limitation of this study is the small complexity of the problem as we only considered able-bodied participants performing simple gestures. We showed that it is sufficient to highlight the cross-subject issue but future research will have to extend this analysis to more complex settings by involving hand gestures with multiple degrees of freedom and an assessment with amputee users.
Finally, to further close the gap between intra-subject and cross-subject estimation, future work will focus on four aspects: i) enlarging the dataset by including additional subjects to enhance the assessment of the approach, ii) extending the search of a common subspace to supervised strategies (e.g., feature selection based on a leave-one-subject-out criterion), iii) making use of more recent and more efficient alignment methods such as those described in \cite{recentSAmethods1} and \cite{recentSAmethods2}, 
and iv) finding a strategy to incorporate the variability of all the source subjects better than naive pooling, which assumes a single source domain.

\subsubsection{Acknowledgements} We gratefully thank all the members of the Laboratory of Neurophysiology and Movement Biomechanics (ULB) for the expertise and equipment they provided us during our data acquisition for this work.

%

\bibliographystyle{splncs04}
\bibliography{bibliography.bib}

\end{document}


\appendix

\section{Data acquisition protocol}
We collected an original dataset of forearm EMG signals (using 8 Cometa Pico electrodes on the right arm) during controlled exercises of 14 participants.
Each subject was asked to keep his right hand open, then hold a specific finger posture (called sign) for at least two seconds before opening back to the resting pose for at least six seconds.
We included 4 signs, each demanding the participant to close every finger except one (thumb, index, middle or little finger). Each sign was repeated 30 times per participant.
We used the motion capture system from an Oculus Quest to detect the sign held by the subject and easily label the samples. This virtual reality device also enabled placing the subject in a controlled environment and showing him which gesture to perform.

To emulate a real-time hand gesture estimation model, a 400ms sliding non-overlapping window was considered to create the samples. We only kept the samples that were recorded during the hold of a sign, not those taken during the gesture that lead to the final posture.
This sampling strategy returns around 150 samples per sign for each participant. 
Each sample contains 400ms of 8 EMG channels with a sampling frequency of 2000Hz.

\begin{figure}
\begin{center}
\includegraphics[width=0.8\textwidth]{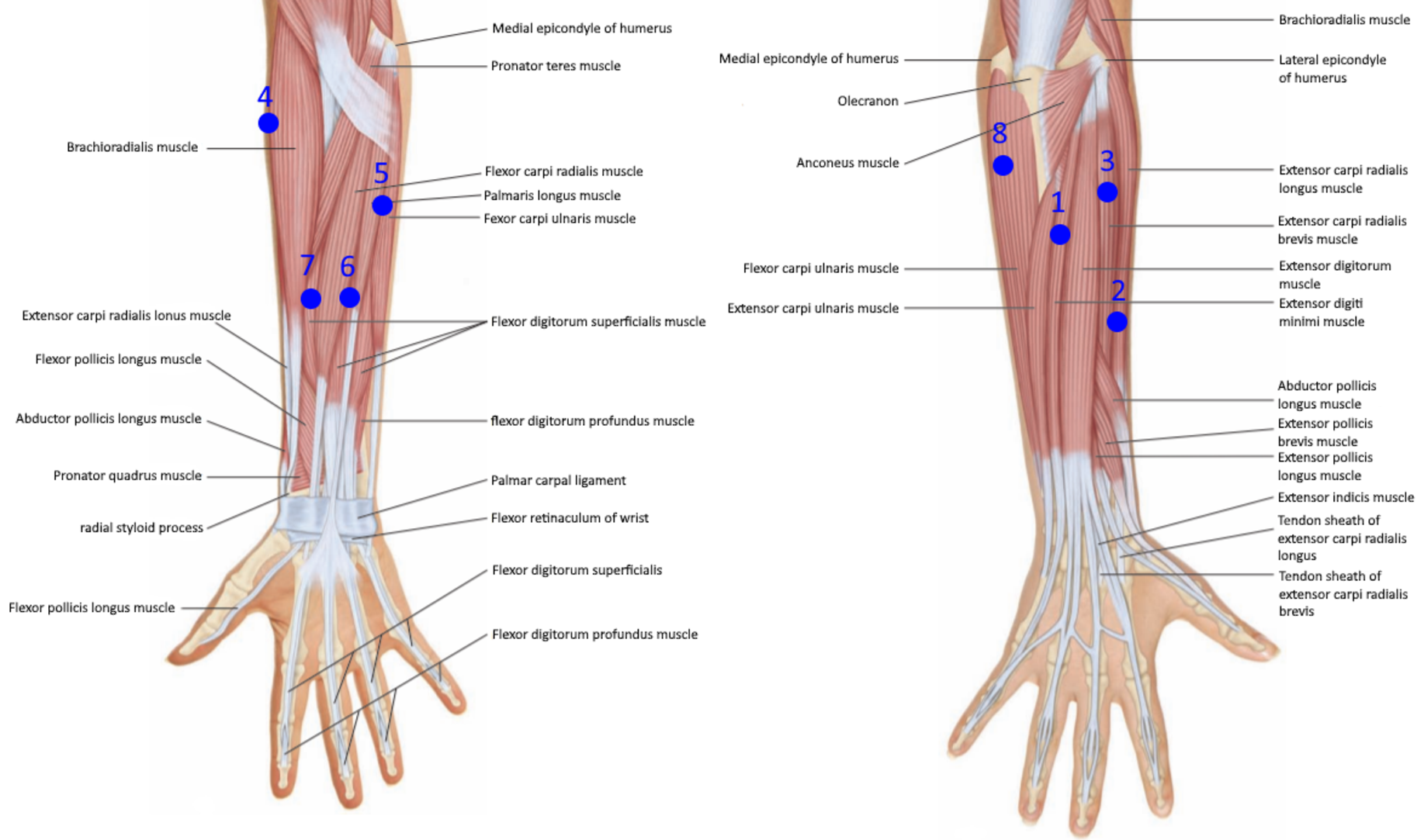}
\caption{Locations of the EMG electrodes on the forearm} \label{fig:EMG_pattern}
\end{center}
\end{figure}

The EMG electrodes were placed using a similar pattern for each participant. This pattern was chosen beforehand from a preliminary experiment performed with a couple of participants. We used electrical stimulation of the forearm so that the flexion and extension of each finger (except the ring finger) are targeted. This pattern is shown in figure \ref{fig:EMG_pattern}.
Even if care was taken to place the electrodes at the same location with each participant, a small location shift likely occurs between each of them.

